\documentclass[%
 reprint,
 amsmath,amssymb,
 aps,prl,]{revtex4-2}

\usepackage{graphicx}
\usepackage{dcolumn}
\usepackage{bm}

\begin{document}

\preprint{APS/123-QED}

\title{Unveiling the Fingerprint of Eccentric Binary Black Hole Mergers}

\author{Hao Wang}
\email{husthaowang@hust.edu.cn}
\affiliation{Department of Astronomy, School of Physics, Huazhong University of Science and Technology, Wuhan 430074, China}

\author{Yuan-Chuan Zou}
\email{zouyc@hust.edu.cn}
\affiliation{Department of Astronomy, School of Physics, Huazhong University of Science and Technology, Wuhan 430074, China}

\author{Qing-Wen Wu}
\email{qwwu@hust.edu.cn}
\affiliation{Department of Astronomy, School of Physics, Huazhong University of Science and Technology, Wuhan 430074, China}

\author{Yu Liu}
\email{yuliu@gzu.edu.cn}
\affiliation{State Key Laboratory of Public Big Data, Guizhou University, Guiyang 550025, China}

\date{\today}

\begin{abstract}
The orbital eccentricity plays a crucial role in shaping the dynamics of binary black hole (BBH) mergers. Remarkably, our recent findings reveal a universal oscillation in essential dynamic quantities: peak luminosity $L_{\text{peak}}$, masses $M_f$, spins $\alpha_f$, and recoil velocity $V_f$ of the final remnant black hole, as the initial eccentricity $e_0$ undergoes variation. In this letter, by leveraging RIT's extensive numerical relativistic simulations of nonspinning eccentric orbital BBH mergers, we not only confirm the universal oscillation in peak amplitudes (including harmonic modes), similar to the oscillations observed in $L_{\text{peak}}$, $M_f$, $\alpha_f$, and $V_f$, but also make the first discovery of a ubiquitous spiral-like internal fine structure that correlates $L_{\text{peak}}$, $M_f$, $\alpha_f$, $V_f$, and peak amplitudes. This distinctive feature, which we term the ``fingerprint" of eccentric orbital BBH mergers, carries important implications for unraveling the intricate dynamics and astrophysics associated with eccentric orbital BBH mergers.
\end{abstract}

\maketitle

\section{Introduction}
Following the groundbreaking detection of the gravitational wave (GW) event GW150914 \cite{LIGOScientific:2016aoc}, the field of GW detection has witnessed a remarkable evolution, transforming into routine practice. Concurrently, numerical relativity (NR), our trusted tool for investigating the dynamics of BBH mergers, has made significant strides in exploring the vast parameter space of BBH systems since its initial breakthrough \cite{Pretorius:2005gq,Campanelli:2005dd,Baker:2005vv}, encompassing configurations spinless systems, spin alignment, spin precession, eccentric orbits and even extreme mass ratios.

Although most of the research on NR and GW detection has focused predominantly on circular orbits, attributed to the circularizing effect of GW radiation \cite{Peters:1963ux,Peters:1964zz}, it is important to recognize various mechanisms through which BBHs can exhibit nonzero eccentricity before their merger. These mechanisms include double-single interactions \cite{Samsing:2013kua,Samsing:2017oij}, double-double interactions \cite{Zevin:2018kzq,Arca-Sedda:2018qgq}, and gravitational capture \cite{Gondan:2020svr,East:2012xq} within dense stellar environments such as globular clusters \cite{OLeary:2005vqo,Rodriguez:2015oxa,Samsing:2017xmd,Rodriguez:2017pec,Rodriguez:2018pss,Park:2017zgj} and galactic nuclei \cite{Gondan:2020svr,Hoang:2017fvh,Gondan:2017wzd,Samsing:2020tda,Tagawa:2020jnc}. Notably, in three-body systems \cite{Naoz:2012bx} involving binary objects orbiting a supermassive black hole, the eccentricity of the inner binary can experience oscillations due to the Kozai-Lidov mechanism \cite{Silsbee:2016djf,Blaes:2002cs,Antognini:2013lpa,Katz:2011hn}. These eccentric BBH systems become detectable once they enter the frequency band of ground-based GW detectors such as LIGO \cite{LIGOScientific:2014qfs}, VIRGO \cite{VIRGO:2014yos} and KAGRA \cite{KAGRA:2018plz}. A notable example is GW190521 \cite{LIGOScientific:2020iuh}, considered a potential BBH merger with a high eccentricity of $e=0.69_{-0.22}^{+0.17}$ \cite{Gayathri:2020coq,Romero-Shaw:2020thy}. With continuous advancements in detector sensitivity, future ground-based GW detectors such as the Einstein Telescope \cite{Punturo:2010zz} or Cosmic Explorer \cite{Reitze:2019iox} are anticipated to observe an increasing number of eccentric BBH mergers.

Over the past decades, several collaborations in NR have conducted extensive simulations of binary compact object mergers, including SXS \cite{Mroue:2013xna,Boyle:2019kee}, RIT \cite{Healy:2017psd,Healy:2019jyf,Healy:2020vre,Healy:2022wdn}, and Georgia Tech. \cite{Jani:2016wkt,Ferguson:2023vta}. These simulations have yielded significant progress in modeling dynamic quantities that hold great astrophysical significance, such as peak luminosity, recoil velocity, remnant mass and spin. Various methods have been employed for modeling these dynamic quantities, including Gaussian Process Regression \cite{Taylor:2020bmj,Varma:2019csw,Varma:2018aht}, post-Newtonian (PN) approaches \cite{Blanchet:2005rj,Racine:2008kj,Buonanno:2007sv}, effective one body methods \cite{Damour:2006tr}, and direct fitting of formulas with NR data \cite{Healy:2016lce,Healy:2014yta,Lousto:2009mf,Hemberger:2013hsa,Lousto:2013wta,Zlochower:2015wga,Lousto:2010xk,Keitel:2016krm}. These approaches enable the modeling of dynamic quantities for both quasi-circular and eccentric orbits, based on initial parameters such as mass ratio and spin.
There have been some investigations of eccentric BBH mergers in recent decades, including studies on the influence of eccentricity on recoil velocity from a PN perspective \cite{Sopuerta:2006et}, the transition from inspiral to plunge in eccentric orbits \cite{Sperhake:2007gu}, orbital circularization for eccentric orbits \cite{Hinder:2007qu}, remnant properties in low eccentricity orbits using NR \cite{Huerta:2019oxn}, kick enhancement caused by eccentricity \cite{Sperhake:2019wwo}, and anomalies in recoil due to eccentricity \cite{Radia:2021hjs}.
Recently, a few articles have explored correlations between the dynamic quantities. For instance, Ref. \cite{Ferguson:2019slp} examined the correlation between the peak amplitude and the remnant spin, while Ref. \cite{Healy:2022wdn} utilized analytical formulas to directly fit NR data for correlations. Furthermore, Ref. \cite{Carullo:2023kvj} obtained a gauge-independent correlation fitted by polynomials, etc. However, these methods did not fully consider the potential fine structures present in NR data.
In our recent work \cite{Wang:2023vka}, incorporating extensive simulation results of BBH mergers in eccentric orbits from RIT's fourth release, we first identified universal oscillations in variations of the aforementioned dynamic quantities as a function of initial eccentricity, which indicate the presence of internal structures in these quantities. In this Letter, we present additional findings of oscillations in peak amplitudes (including higher-order harmonic modes). By summarizing the correlations between quantities such as peak luminosity $L_{\text{peak}}$ (maximum value of dimensionalized radiation energy \cite{Healy:2016lce}), masses $M_f$, spins $\alpha_f$, recoil velocity $V_f$ (magnitude of dimensional recoil velocity), and peak amplitudes (maximum amplitude values)   $\mathcal{A}_{22,\text{peak}}$, $\mathcal{A}_{32,\text{peak}}$, $\mathcal{A}_{44,\text{peak}}$, we unveil, for the first time, the existence of a spiral-like internal fine structure in these correlations. We refer to this structure as the ``fingerprint" of eccentric orbital BBH mergers according to its characteristics. 

Throughout this letter, we adopt geometric units where $G=c=1$. The component masses of BBH are represented as $m_1$ and $m_2$, while the total mass is denoted by $M=m_1+m_2$. For simplicity, we set the total mass $M$ at unity (sometimes explicitly writing it for clarity). The mass ratio $q$ is defined as $q = m_1/m_2$, and $m_1<m_2$. 

\section{Method}
We utilize NR simulations of BBH systems in eccentric orbits obtained from the Rochester Institute of Technology (RIT) catalog \cite{RITBBH}. These simulations were conducted using the LazEv code \cite{Zlochower:2005bj}, implemented within the Einstein Toolkit \cite{Loffler:2011ay} alongside the CACTUS/CARPET infrastructure \cite{Schnetter:2003rb}. The LazEv code employs the moving puncture approach \cite{Campanelli:2005dd} and utilizes the BSSNOK formalism for evolution systems \cite{Nakamura:1987zz,Shibata:1995we,Baumgarte:1998te}.
In the initial stages, RIT employed AHFinderDirect \cite{Thornburg:2003sf} to locate apparent horizons and employed the isolated horizon algorithm to measure the amplitude of the horizon spins \cite{Campanelli:2006fy}, denoted as $S_H$. Subsequently, they calculated the horizon mass using the Christodoulou formula: $m_H=\sqrt{m_{\mathrm{irr}}^2+S_H^2 /\left(4 m_{\mathrm{irr}}^2\right)}$. Here, $m_{\mathrm{irr}}$ represents the irreducible mass, defined as $m_{\mathrm{irr}}=\sqrt{A_H /(16 \pi)}$, where $A_H$ corresponds to the surface area of the horizon \cite{Campanelli:2006fy}.

In generating the initial data, RIT adopts the puncture approach \cite{Brandt:1997tf} in combination with the TwoPunctures code \cite{Ansorg:2004ds}. To enable continuous-eccentricity simulations, RIT initially employs PN techniques, as outlined in Ref. \cite{Healy:2017zqj}, to generate initial data for quasi-circular orbits. Subsequently, by introducing a new parameter $\epsilon$ within the range of 0 to 1, the tangential linear momentum is modified according to $p_t=p_{t,qc}(1-\epsilon)$. Within this framework, the initial positions of the BBHs are fixed at the apocenter, and the eccentricity of the orbit gradually increases throughout the simulations, spanning from the quasi-circular orbit ($e_0=0$) to the head-on collision limit ($e_0=1$). The initial eccentricity $e_0$ of the orbit can be approximated by $e_0=2\epsilon-\epsilon^2$, offering a second-order approximation in terms of $\epsilon$ that accurately captures the limits of $e_0=0$ and $e_0=1$ at $\epsilon=0$ and $\epsilon=1$, respectively \cite{Healy:2022wdn}.

The RIT catalog provides a comprehensive dataset comprising waveform data and accompanying metadata \cite{RITBBH}. In our research, we utilize the gravitational wave strain $h$ as the waveform data, which can be obtained as the harmonic mode $rh=\sum_{l, m}rh_{l m}{ }_{-2} Y_{l, m}(\theta, \phi)$, where $r$ represents the extracted radius, and ${ }_{-2} Y_{l, m}(\theta, \phi)$ denotes the spin-weighted spherical harmonic functions. The harmonic mode $h_{l m}$ can be decomposed into amplitude and phase as $h_{lm}=\mathcal{A}_{lm}(t) \exp \left[-i \Phi_{lm}(t)\right]$.
The metadata within the catalog provides essential information regarding the initial data of the simulations, which includes details such as the mass ratio, initial distance, initial linear momentum, and more. Additionally, the metadata contain significant simulation results, such as the peak luminosity $L_{\text{peak}}$, masses $M_f$, spins $\alpha_f$, and recoil velocity $V_f$ of the final remnant black hole.
In all simulations conducted by RIT, it has been ensured that the waveforms, at the resolutions provided in the catalog, have achieved convergence up to fourth order with resolution \cite{Healy:2022wdn}. The evaluation of quantities related to the black hole horizon, such as the final masses $M_f$ and spins $\alpha_f$ of the remnant, yields errors of the order of 0.1\% using the isolated horizon algorithm \cite{Campanelli:2006fy}. Furthermore, the radiative computed quantities, including the recoil velocities $V_f$ and the peak luminosities $L_{\text{peak}}$, are evaluated with a typical error of 5\% \cite{Healy:2022wdn}. Therefore, all of the data provided in the catalog meet the precision requirements for research purposes.

\begin{figure}[htbp!]
\centering
\includegraphics[width=8cm,height=5cm]{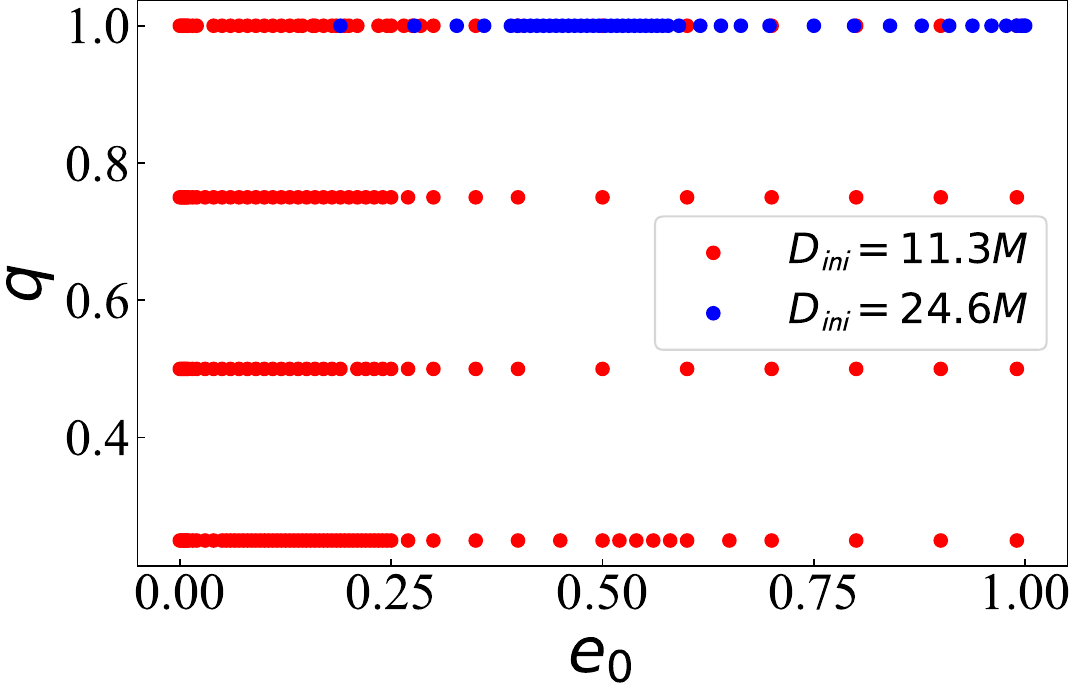}
\caption{\label{FIG:1}Parameter space of the eccentric nonspinning BBH NR simulations used in our study. There are 192 sets simulations with $D_{\text{ini}}=11.3M$ and $q=1$ (43 sets), $q=0.25$ (67 sets), $q=0.5$ (41 sets), $q=0.75$ (41 sets), and 48 sets simulations with $D_{\text{ini}}=24.6M$ and $q=1$.}
\end {figure}

In this letter, our focus is solely on the nonspinning configuration, as it provides a larger dataset compared to spin alignment and spin precession simulations \cite{Wang:2023vka}. In addition to the previously mentioned dynamic quantities $L_{\text{peak}}$, $M_f$, $\alpha_f$, and $V_f$, we also introduce the peak amplitudes of different harmonic modes, specifically $\mathcal{A}_{22,\text{peak}}$, $\mathcal{A}_{32,\text{peak}}$, and $\mathcal{A}_{44,\text{peak}}$. We have chosen not to investigate other harmonic modes and peak frequencies due to issues with the former in RIT's catalog (although we speculate that other harmonics behave similarly) and the fact that the latter does not accurately represent the merger structure of BBH systems. Our analysis focuses on characterizing the variations of the peak amplitudes as functions of $e_0$ and examining the correlations between the dynamic quantities and the peak amplitudes.
Fig. \ref{FIG:1} presents the parameter space of the eccentric nonspinning BBH NR simulations used in our study. Specifically, we analyze a total of 192 data sets for initial coordinate separation $D_{\text{ini}}=11.3M$ and mass ratios $q=1, 0.75, 0.5, 0.25$, along with 48 data sets for $D_{\text{ini}}=24.6M$ and mass ratio $q=1$. Due to the limited availability of simulated data points, nonspinning simulations for other mass ratios, as well as simulations involving spin alignment and spin precession, cannot provide sufficient information for our research \cite{Wang:2023vka}.

\section{Results}
\begin{figure}[htbp!]
\centering
\includegraphics[width=8cm,height=5cm]{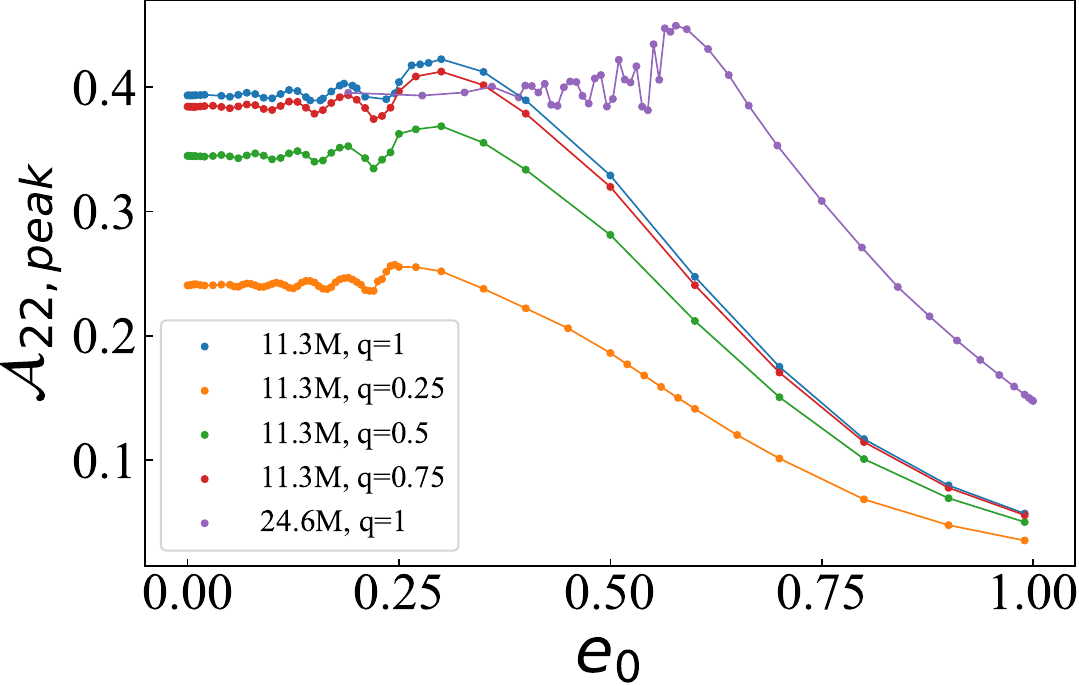}
\caption{\label{FIG:2}Variations of 2-2 mode peak amplitude $\mathcal{A}_{22,\text{peak}}$ as a function of initial eccentricity $e_0$ at $D_{\text{ini}}=11.3M$ and $D_{\text{ini}}=24.6M$ for nonspinning configuration with different mass ratios.}
\end {figure}

\begin{figure*}[htbp!]
\centering
\includegraphics[width=16cm,height=10cm]{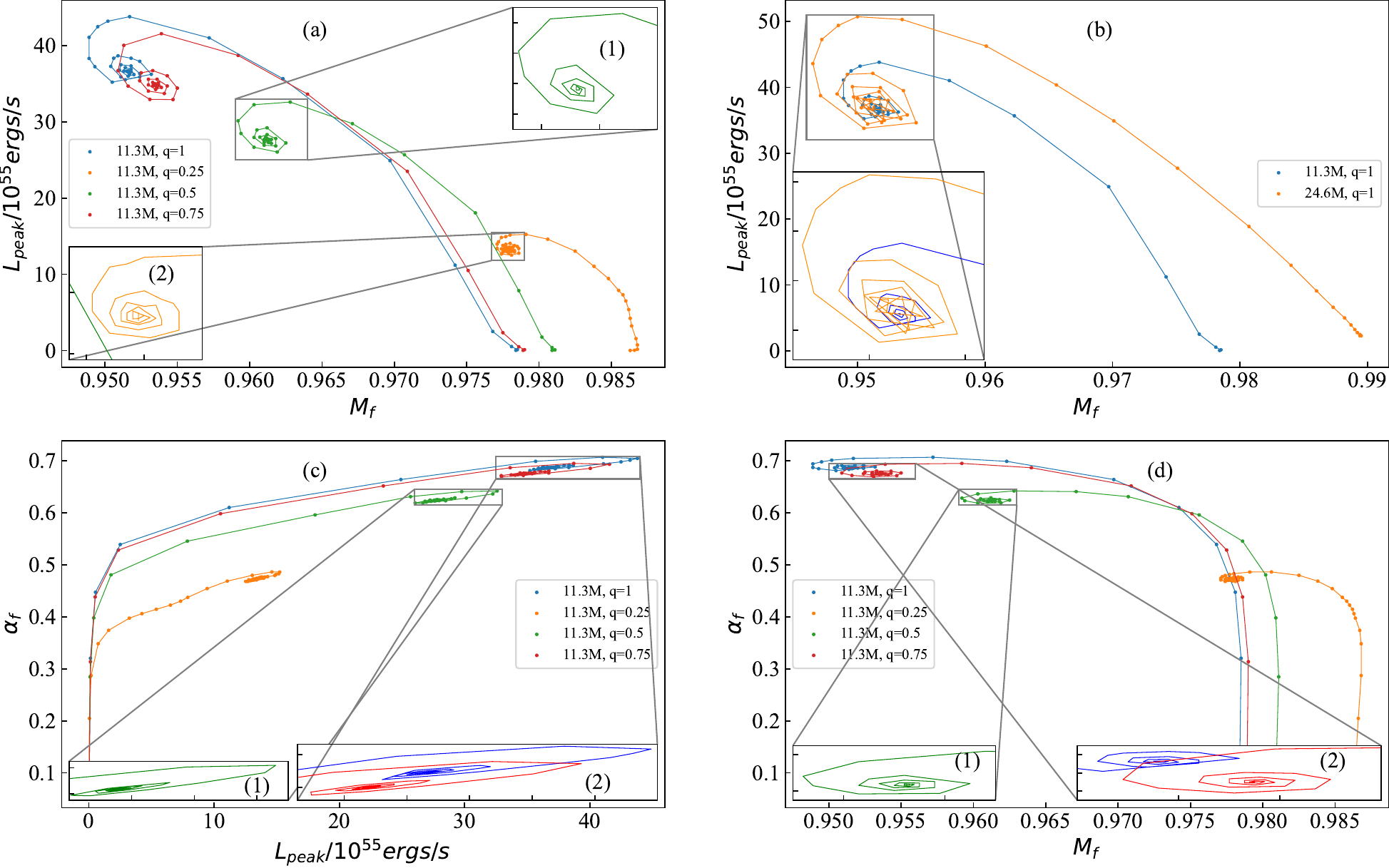}
\caption{\label{FIG:3}Correlations between quantities $L_{\text{peak}}$, $M_f$, and $\alpha_f$, for four mass ratios and two initial coordinate separations $D_{\text{ini}}=11.3M$ and $D_{\text{ini}}=24.6M$. The ``fingerprints" are drawn in insets as partial enlargement for details.}
\end {figure*}

In our previous study, described in Ref. \cite{Wang:2023vka}, we investigate the universal oscillations observed in dynamical quantities $L_{\text{peak}}$, $M_f$, $\alpha_f$, and $V_f$ as a function of $e_0$. Supplementary to these findings, Fig. \ref{FIG:2} presents the relationship between the peak amplitude of the 2-2 mode $\mathcal{A}_{22,\text{peak}}$ and $e_0$ for both $D_{\text{ini}}=11.3M$ and $D_{\text{ini}}=24.6M$ (for $\mathcal{A}_{32,\text{peak}}$, $\mathcal{A}_{44,\text{peak}}$, refer to Fig. \ref{FIG:s4} of supplementary materials). Each point in Fig. \ref{FIG:2} corresponds to a simulation result. Fig. \ref{FIG:2} reveal the same universal oscillatory behavior observed in the peak amplitudes. While these oscillations share similarities with peak luminosity $L_{\text{peak}}$ oscillations described in Ref. \cite{Wang:2023vka}, they also demonstrate distinct characteristics influenced by mass ratio $q$ when considering higher harmonic modes.
In our previous work \cite{Wang:2023vka}, we proposed that these peculiar oscillations may arise from orbital transitions based on the integer orbital cycles $N_{\text {waves }}=\frac{\Delta \Phi}{4\pi}$ of GW. Here, the phase difference $\Delta \Phi$ is calculated as $\Delta \Phi=\Phi\left(t_{\mathrm{merger}}\right)-\Phi\left(t_0+t_{\mathrm{relax}}\right)$ from the 2-2 mode of GW, where $t_{\mathrm{merger}}$, $t_0$, and $t_{\mathrm{relax}}$ represent the merger moment, initial moment and duration of the junk radiation, respectively. These fine oscillatory structures can only be discerned when there is a sufficient number of data points from eccentric numerical simulations \cite{Wang:2023vka}. In Fig. \ref{FIG:s5} of  supplementary material, we provide further evidence to emphasize their relevance by illustrating the relationship between peak amplitudes and the integer $N_{\text {waves }}$ as described in Ref. \cite{Wang:2023vka}.
The shift and enhancement of the oscillations for higher $e_0$ for $D_{\text{ini}}=24.6M$ compared to $D_{\text{ini}}=11.3M$, along with the correspondence between integer $N_{\text {waves }}$ and the peak and valley points, indicates that these oscillations originate from strong field dynamics, as highlighted in Ref. \cite{Wang:2023vka}. Consequently, we aim to explore the impact of this strong field effect due to the existence of eccentricity on the correlations between the dynamic quantities $L_{\text{peak}}$, $M_f$, $\alpha_f$ and $V_f$, and peak amplitudes $\mathcal{A}_{22,\text{peak}}$, $\mathcal{A}_{32,\text{peak}}$, $\mathcal{A}_{44,\text{peak}}$. Fig. \ref{FIG:3} showcases three representative correlations between the aforementioned dynamic quantities and peak amplitudes. To maintain brevity, we include the remaining 18 correlations in Figs. \ref{FIG:s6}, \ref{FIG:s7} and \ref{FIG:s8} of supplementary material.

In Fig. \ref{FIG:3}, we present the functional relationships between the quantities $L_{\text{peak}}$, $M_f$, and $\alpha_f$ for four mass ratios and two initial coordinate separations. Partial enlargements are included as insets for a detailed examination. Taking panels (a) and (b) in Fig. \ref{FIG:3} as examples, we observe a captivating spiral structure in the curves. The center of the spiral corresponds to the quasi-circular orbit counterpart ($e_0=0$) of this family of eccentric orbital BBH merger simulations. As the initial eccentricity $e_0$ gradually increases, the spiral rotates outwards. When the spiral reaches the transition point from inspiral to plunge (the maximum or minimum value of oscillations, see Ref. \cite{Wang:2023vka}), it terminates, and the curve gradually reaches the specific value corresponding to the maximum initial eccentricity (i.e., $e_0=1$ for the head-on collision limit).
The insets (1) and (2) in panel (a) represent the mass ratios $q=0.5$ (40 data points) and $q=0.25$ (67 data points), respectively, for $D_{\text{ini}}=11.3M$. From panel (a) of Fig. \ref{FIG:3}, we observe that a smaller mass ratio $q$ corresponds to a smaller spiral, originating weaker oscillations of the corresponding quantity with changes in $e_0$, as observed in Ref. \cite{Wang:2023vka}. The right panel (b) corresponds to the case of the same mass ratio $q=1$ and different initial separations $D_{\text{ini}}=11.3M$ and $D_{\text{ini}}=24.6M$. We can discern that a larger $D_{\text{ini}}$ corresponds to a larger spiral, originating stronger oscillations of the corresponding quantity with changes in $e_0$.
In panels (c) and (d), the spiral becomes more elliptical; however, we still observe a similar spiral structure. The elliptical shape is due to the weaker oscillation of $\alpha_f$ compared to $L_{\text{peak}}$ and $M_f$ (see Ref. \cite{Wang:2023vka}).  In Ref. \cite{Wang:2023vka}, we obtained irregular oscillations in $V_f$ and provided an explanation. When any other quantity is combined with $V_f$, an irregular spiral pattern emerges (please refer to the supplementary material). These curves of $V_f$ exhibit various irregular behaviors and overlaps, distinguishing them from panels (a), (b), (c), (d). However, it is important to acknowledge that, despite their irregularity, their internal structure remains intact.
Based on the supplementary material and Fig. \ref{FIG:3}, we can conclude that a ubiquitous spiral structure is present in the correlations between the dynamic quantities $L_{\text{peak}}$, $M_f$, $\alpha_f$, $V_f$, and peak amplitudes. The position and size of a spiral are characterized by the mass ratio $q$ and initial coordinate separation $D_{\text{ini}}$. Two types of spiral structures are identified: one is irregular, corresponding to the irregularly oscillating $V_f$, while the other is regular and not combined with $V_f$. Due to the spiral structure observed in these correlations, we refer to the prevalent correlations in eccentric orbital BBH mergers as ``fingerprint", which uniquely represents a family of eccentric BBH merger simulations, analogous to how a fingerprint represents an individual.

In some cases, the orbital transition points of the integer $N_{\text {waves}}$ may not precisely align with the peaks and valleys of oscillation. Several factors contribute to this discrepancy, including errors and pericenter precession \cite{Wang:2023vka}. To investigate the relationship between the spirals and the peaks and valleys of oscillation, we mark the points corresponding to the peaks and valleys in the relationship between the dynamic quantities $L_{\text{peak}}$, $M_f$, $\alpha_f$, and $e_0$ on the spirals in Fig. \ref{FIG:s9} of supplementary material. Notably, we find that the peaks and valleys of the oscillation are positioned on opposite sides of the spiral and are concentrated near a straight line. This phenomenon suggests that the spiral structure arises from the oscillations of $L_{\text{peak}}$, $M_f$, and $\alpha_f$, which in turn originate from the orbital transitions of the periodic process.

\section{Discussion}
Previously, many studies have approached modeling dynamic quantities by directly fitting scatter points using methods such as polynomials. However, the existence of internal fine structures within these dynamic quantities, particularly in relation to eccentricity, was not recognized. As we discussed previously, these internal structures originate from the oscillations of dynamic quantities as a function of eccentricity, which are driven by the strong-field dynamics of eccentric BBH mergers, specifically, orbital transitions that correspond to different orbital cycle numbers. This unique characteristic distinguishes BBH mergers in eccentric orbits from those in quasi-circular orbits, where the spiral structure dissipates and returns to the central point.

In summary, by leveraging RIT's extensive numerical relativity simulations of nonspinning eccentric orbital binary black hole mergers, we make the first discovery of a universal spiral-like internal fine structure in the correlations between various quantities, including $L_{\text{peak}}$, $M_f$, $\alpha_f$, and peak amplitudes such as $\mathcal{A}_{22,\text{peak}}$, $\mathcal{A}_{32,\text{peak}}$, and $\mathcal{A}_{44,\text{peak}}$, arising due to eccentricity. This spiral structure presents a fresh perspective on the investigation of eccentric orbital BBH mergers and their strong-field dynamics. Furthermore, it holds significant implications for the study of eccentric orbital BBH mergers within dense stellar environments like globular star clusters and galactic nuclei \cite{Samsing:2017oij,Samsing:2017xmd}. Moreover, it enables a more precise understanding and modeling of various astrophysical quantities, including remnant black hole mass, spin and recoil velocity.

It is important to note that our examination of nonspinning eccentric BBHs represents only a subset of the parameter space explored by NR simulations. In the future, we anticipate the generation of additional simulations that encompass spin-aligned and spin-precessing configurations, which will offer crucial insights into the interplay between spin and eccentricity in strong-field dynamics.

\begin{acknowledgments}
The authors are very grateful to the RIT collaboration for the numerical simulation of eccentric BBH mergers, and thanks to Yan-Fang Huang, Zhou-Jian Cao for their helpful discussions. The computation is partially completed in the HPC Platform of Huazhong University of Science and Technology. The languages was polished by ChatGPT during the revision of the draft. This work is supported by the National Key R\&D Program of China (2021YFA0718504).
\end{acknowledgments}


\bibliography{ref}

{\newpage
\section{Supplementary material}

\renewcommand{\theequation}{S\arabic{equation}}
\renewcommand{\thefigure}{S\arabic{figure}}
\renewcommand{\bibnumfmt}[1]{[S#1]}
\renewcommand{\citenumfont}[1]{S#1}

\preprint{APS/123-QED}

\title{Supplemental materials}

\date{\today}

\maketitle

\section{Oscillations in peak amplitudes}

In this section, in Fig. \ref{FIG:s4} we show the variations of higher order harmonics peak amplitudes, $\mathcal{A}_{32,\text{peak}}$ (panel (a)), $\mathcal{A}_{44,\text{peak}}$ (panel (b)) as a function of the initial eccentricity $e_0$ at the initial coordinate separation of $D_{\text{ini}}=11.3M$ and $D_{\text{ini}}=24.6M$ for nonspinning configuration with different mass ratio $q=1, 0.75, 0.5, 0.25$. We can see that there is a universal oscillation in higher order harmonics peak amplitudes similar to $L_{\text{peak}}$, masses $M_f$, spins $\alpha_f$, recoil velocity $V_f$, and $\mathcal{A}_{22,\text{peak}}$ \cite{Wang:2023vka}. 

\setcounter{figure}{0}
\begin{figure*}[htbp!]
\centering
\includegraphics[width=12cm,height=5cm]{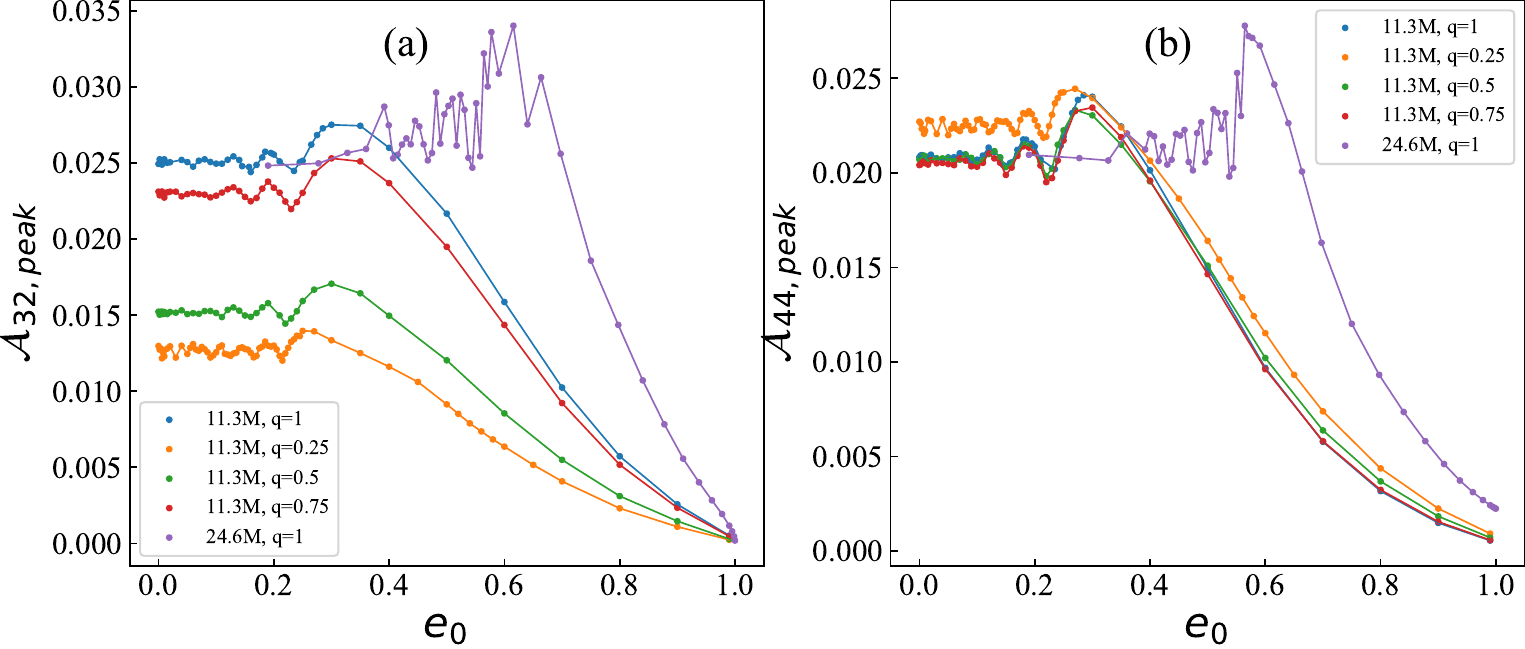}
\caption{\label{FIG:s4}Variations of higher order harmonics peak amplitudes $\mathcal{A}_{32,\text{peak}}$ (panel (a)), $\mathcal{A}_{44,\text{peak}}$ (panel (b)) as a function of the initial eccentricity $e_0$ at the initial coordinate separation of $D_{\text{ini}}=11.3M$ and $D_{\text{ini}}=24.6M$ for nonspinning configuration with different mass ratio $q=1, 0.75, 0.5, 0.25$.} 
\end {figure*}

\section{Integer orbital cycles in peak amplitudes}
In this section, we follow the ideas of Ref. \cite{Wang:2023vka}
to consider the orbital transition. The number of orbital cycles $N$ can be determined through the phase of the gravitational waveform. In our analysis, we specifically focus on the 2-2 mode. To calculate the phase difference, we evaluate the expression:
\begin{equation}\label{eq:1}
\Delta \Phi=\Phi\left(t_{\mathrm{merger}}\right)-\Phi\left(t_0+t_{\mathrm{relax}}\right).
\end{equation}
where $t_{\mathrm{merger}}$ represents the time of BBH merger, $t_0$ denotes the initial moment of the waveform, and $t_{\mathrm{relax}}$ signifies the time required to the transition from the initial moment to a physically stable state. 
The number of orbital cycles accomplished by the BBH system can be obtained as:
\begin{equation}\label{eq:s2}
N_{\text {waves }}=\frac{\Delta \Phi}{4 \pi}.
\end{equation}
Here, we divide the phase difference $\Delta \Phi$ by $4 \pi$ since the waveform phase of 2-2 mode is twice that of the orbital phase. Fig. \ref{FIG:s5} displays the relationship between the integer orbital cycle number $N_{\text{waves}}$ and peak amplitudes $\mathcal{A}_{22,\text{peak}}$ (panel (a)), $\mathcal{A}_{32,\text{peak}}$ (panel (b)), $\mathcal{A}_{44,\text{peak}}$ (panel (c)) at initial coordinate separations $D_{\text{ini}}=11.3M$ and $D_{\text{ini}}=24.6M$ for nonspinning configuration with different mass ratio $q=1, 0.75, 0.5, 0.25$. These points, denoted by red ``x" markers, correspond to either an integer multiple or are in close proximity to an integer multiple of the orbital cycles. Moving from right to left, each red ``x" corresponds to successive orbital cycles, starting from cycle 1 and continuing indefinitely.

\begin{figure*}[htbp!]
\centering
\includegraphics[width=16cm,height=5cm]{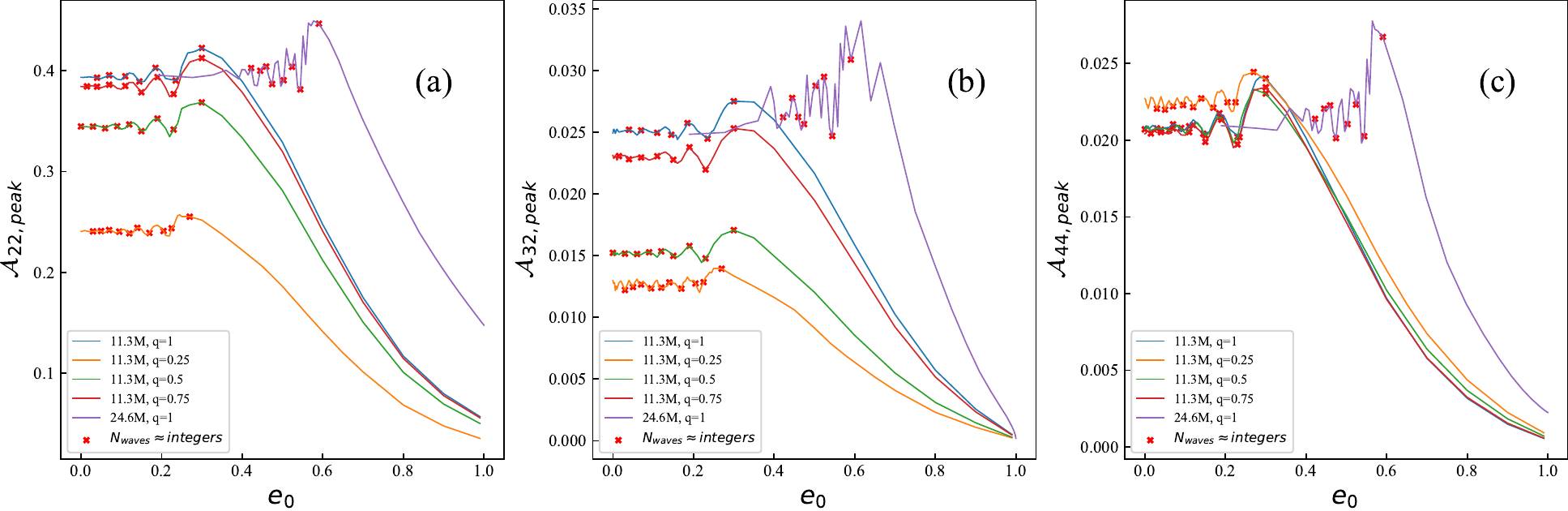}
\caption{\label{FIG:s5}Relationship between the integer orbital cycle number $N_{\text{waves}}$ and peak amplitudes $\mathcal{A}_{22,\text{peak}}$ (panel (a)), $\mathcal{A}_{32,\text{peak}}$ (panel (b)), $\mathcal{A}_{44,\text{peak}}$ (panel (c)) at initial coordinate separations $D_{\text{ini}}=11.3M$ and $D_{\text{ini}}=24.6M$ for nonspinning configuration with different mass ratio $q=1, 0.75, 0.5, 0.25$. These points, denoted by red ``x" markers, correspond to either an integer multiple or are in close proximity to an integer multiple of the orbital cycles. Moving from right to left, each red ``x" corresponds to successive orbital cycles, starting from cycle 1 and continuing indefinitely.}
\end {figure*}

\section{Other correlations}
In this section, we show the other 18  correlations between dynamic quantities $L_{\text{peak}}$, masses $M_f$, spins $\alpha_f$, and recoil velocity $V_f$ and peak amplitudes $\mathcal{A}_{22,\text{peak}}$ , $\mathcal{A}_{32,\text{peak}}$, $\mathcal{A}_{44,\text{peak}}$ in Fig. \ref{FIG:s6}, Fig. \ref{FIG:s7} and Fig. \ref{FIG:s8}.

\setcounter{figure}{2}
\begin{figure*}[htbp!]
\centering
\includegraphics[width=15cm,height=22cm]{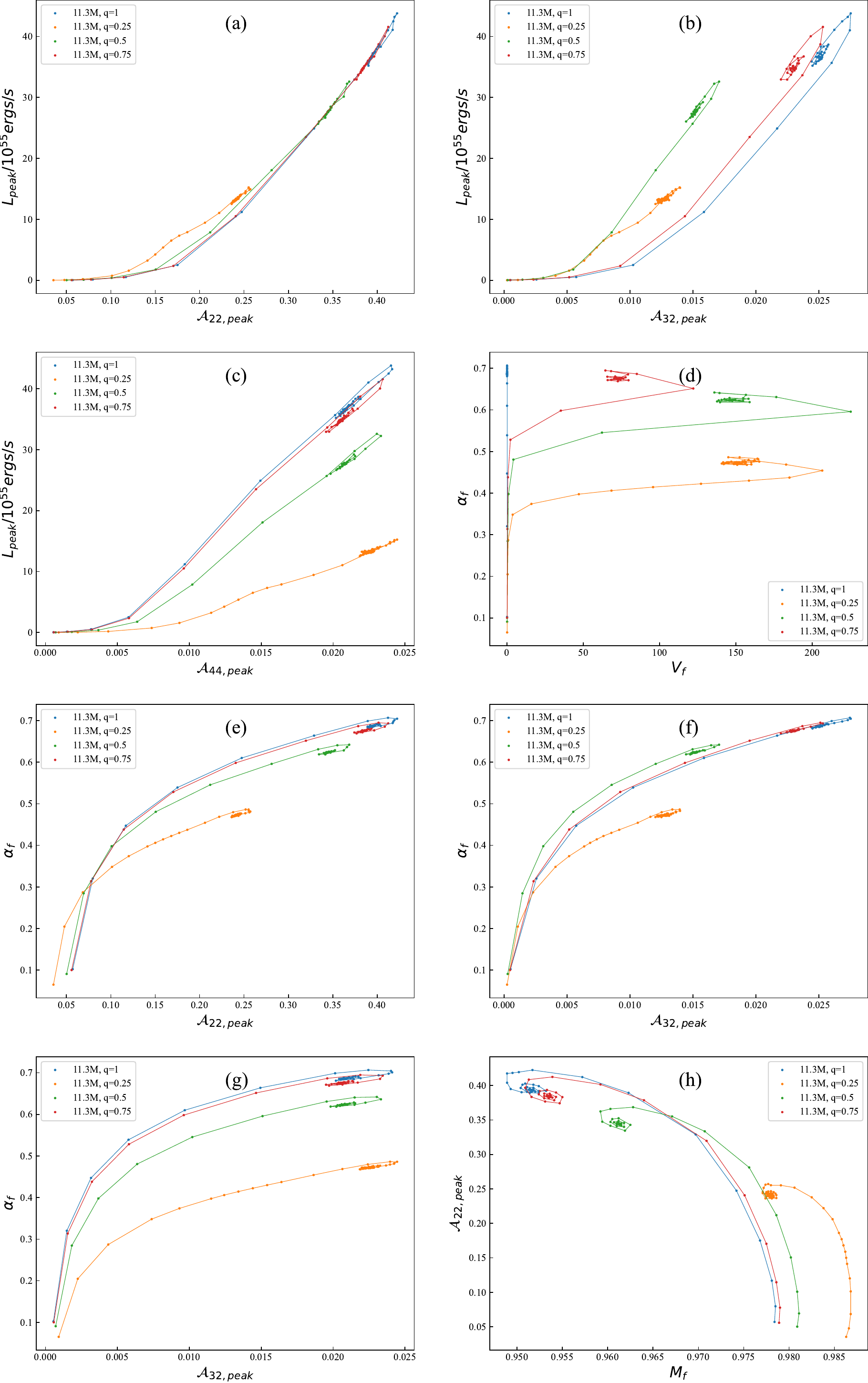}
\caption{\label{FIG:s6}Correlations between dynamic quantities $L_{\text{peak}}$, masses $M_f$, spins $\alpha_f$, and recoil velocity $V_f$ and peak amplitudes $\mathcal{A}_{22,\text{peak}}$ , $\mathcal{A}_{32,\text{peak}}$, $\mathcal{A}_{44,\text{peak}}$.}
\end {figure*}

\begin{figure*}[htbp!]
\centering
\includegraphics[width=15cm,height=22cm]{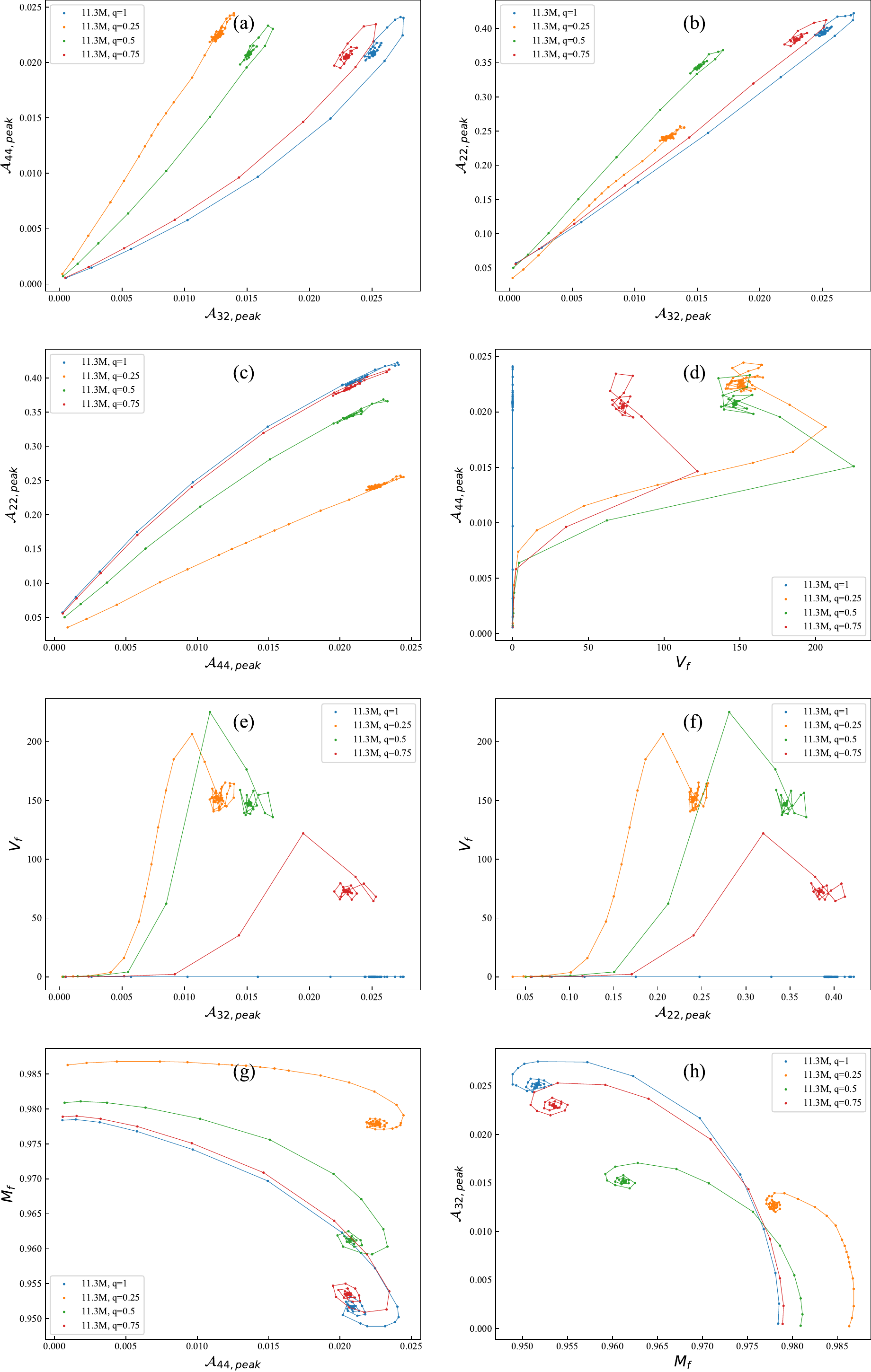}
\caption{\label{FIG:s7}Correlations between dynamic quantities $L_{\text{peak}}$, masses $M_f$, spins $\alpha_f$, and recoil velocity $V_f$ and peak amplitudes $\mathcal{A}_{22,\text{peak}}$ , $\mathcal{A}_{32,\text{peak}}$, $\mathcal{A}_{44,\text{peak}}$.}
\end {figure*}

\begin{figure*}[htbp!]
\centering
\includegraphics[width=15cm,height=5.5cm]{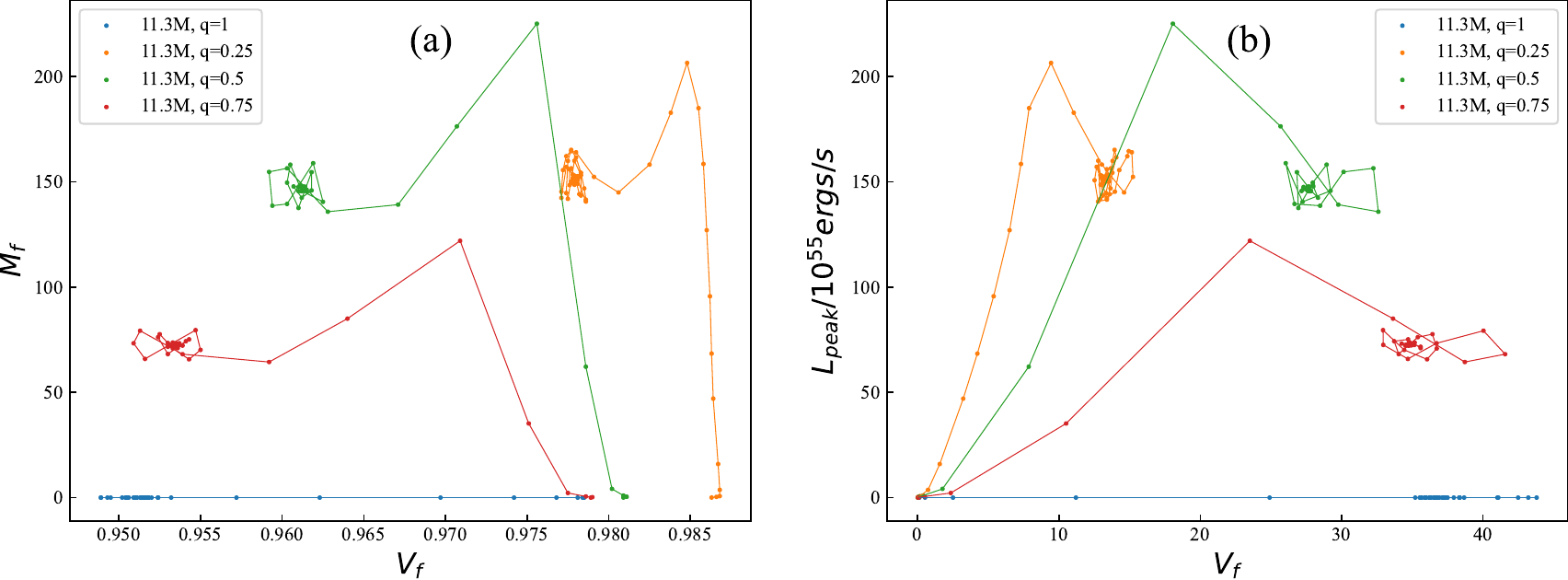}
\caption{\label{FIG:s8}Correlations between dynamic quantities $L_{\text{peak}}$, masses $M_f$, spins $\alpha_f$, and recoil velocity $V_f$ and peak amplitudes $\mathcal{A}_{22,\text{peak}}$ , $\mathcal{A}_{32,\text{peak}}$, $\mathcal{A}_{44,\text{peak}}$.}
\end {figure*}

\section{The origin of fingerprints}
In this section, we select two typical cases to analyze the relationship between oscillation and this spiral structure. Other situations are the same as it. We mark the positions of the peaks and valleys of oscillations of dynamic quantities $L_{\text{peak}}$, masses $M_f$, spins $\alpha_f$ in the spiral structure in Fig. \ref{FIG:s9}. Notably, we find that the peaks and valleys of oscillation are positioned on opposite sides of the spiral and
are concentrated near a straight line. For $M_f$ and $L_{\text{peak}}$,
there appears to be an approximate phase difference (see panel (a), (c), (e), (g) of Fig. \ref{FIG:s9}), while $L_{\text{peak}}$ and $\alpha_f$ nearly
coincide (see panel (b), (d), (f), (h) of Fig. \ref{FIG:s9}).

\begin{figure*}[htbp!]
\centering
\includegraphics[width=15cm,height=22cm]{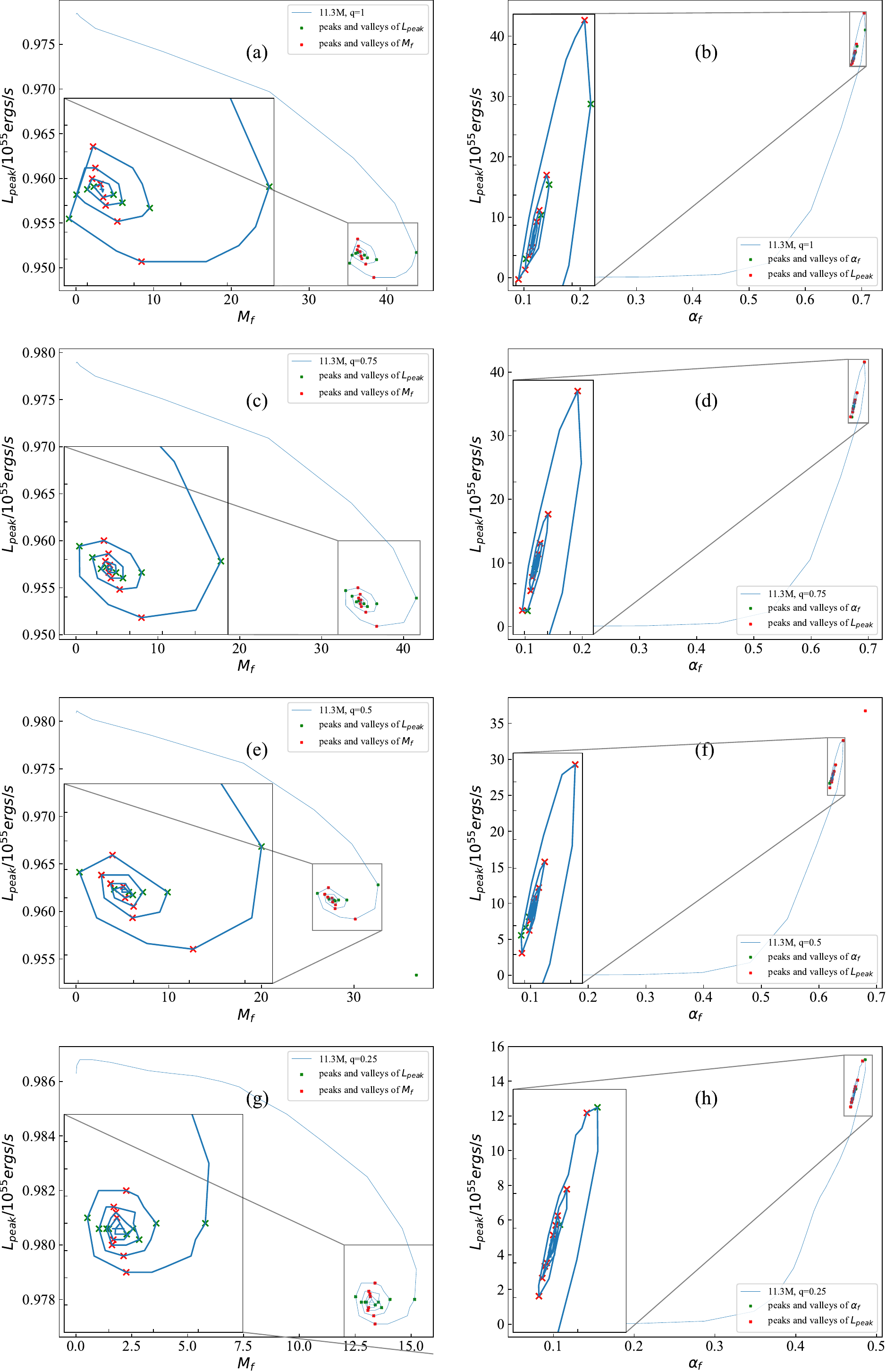}
\caption{\label{FIG:s9} Points corresponding to the peaks and valleys of oscillations in the relationship between dynamic quantities
$L_{\text{peak}}$, masses $M_f$, spins $\alpha_f$ and $e_0$ marked with red ``x''.}
\end {figure*}

}

\end{document}